\documentclass[prl,twocolumn,showpacs,superscriptaddress,
preprintnumbers,amsmath,amssymb,tightenlines,floatfix]{revtex4}
\usepackage{graphicx}
\newcommand{\beq}{\begin{equation}}
\newcommand{\eeq}{\end{equation}}
\newcommand{\bea}{\begin{eqnarray}}
\newcommand{\eea}{\end{eqnarray}}
\newcommand{\ba}{\begin{array}}
\newcommand{\ea}{\end{array}}
\newcommand{\bc}{\begin{center}}
\newcommand{\ec}{\end{center}}
\newcommand{\lsimeq}{\alt}

\newcommand{\bml}{\begin{subequations}}
\newcommand{\eml}{\end{subequations}}
\newcommand{\commentout}[1]{{}}

\newcommand{\half}{\hbox{$\frac{1}{2}$}}

\newcommand{\etal} {{\it et al.\/}}

\newcommand{\comment}[1]{{}}
\begin{document}
\title{Comment on ``Stimulated Raman adiabatic passage from an atomic to
a molecular Bose-Einstein condensate"}

\author{Matt Mackie}
\altaffiliation [Currently also at ]{Department of Physics, University of
Turku, FIN-20014 Turun yliopisto, Finland.}
\affiliation{Helsinki Institute of Physics, PL 64, FIN-00014
Helsingin yliopisto, Finland}
\author{Anssi Collin}
\affiliation{Helsinki Institute of Physics, PL 64, FIN-00014
Helsingin yliopisto, Finland}
\author{Juha Javanainen}
\affiliation{Department of Physics, University of Connecticut,
Storrs, Connecticut, 06269-3046}
\affiliation{Optics and Molecular Materials, PL 2200,
FIN-02015 Helsinki University of Technology, Finland}

\date{\today}

\begin{abstract}
Collective two-color photoassociation of a freely-interacting
$^{87}$Rb Bose-Einstein condensate is theoretically examined, focusing on
stimulated Raman adiabatic passage (STIRAP) from an atomic to a stable
molecular condensate. In particular, Drummond~\etal~{[\pra
{\bf 65}, 063619 (2002)]} have predicted that particle-particle
interactions can limit the efficiency of collective atom-molecule STIRAP,
and that optimizing the laser parameters can partially overcome this
limitation. We suggest that the molecular conversion efficiency can be
further improved by treating the initial condensate density as an
optimization parameter.
\end{abstract}


\maketitle

Stimulated Raman adiabatic passage (STIRAP) from an
atomic to a molecular condensate\cite{MAC00} in the presence of
particle-particle interactions was recently
investigated~\cite{HOP01,DRU02}. In particular,
Drummond {\it et al.}~\cite{DRU02} predict that, for a $^{87}$Rb
Bose-Einstein condensate (BEC) of typical density  ($\rho\sim
10^{14}\,{\rm cm}^3$), the STIRAP conversion efficiency can be limited in
practice by two-photon dephasing caused by particle interactions,
and that said limitation can be partially overcome by optimizing the
laser parameters. The purpose of this Comment is to suggest that the role
of particle-particle interactions can be further downplayed, and the
conversion efficiency improved, by treating the
initial condensate density as an additional optimization parameter.

The mean-field equations for collective two-color
photoassociation of a freely-interacting gas can be written
\bml
\bea
i\dot{a} &=&
  \left(\half\Delta+\Lambda_{aa}|a|^2+\Lambda_{ag}|g|^2\right)a
    - \chi a^* b, \\
i\dot{b} &=& \left(\delta-\half i\gamma_s \right) b
  -\half\left(\chi aa + \Omega g\right), \\
i\dot{g} &=& \left(\Lambda_{ag} |a|^2 + \Lambda_{gg} |g|^2\right) g
  -\half\Omega b,
\eea
\eml
where the complex amplitudes $a$, $b$, and $g$ represent the
respective atomic, excited-molecular, and stable-molecular condensates.
The laser-matter interactions that drive the
respective atom-molecule and molecule-molecule transitions are
$\chi(t)=\chi_0\exp[-\left(t-D_1\right)^2/T^2]$ and
$\Omega(t)=\Omega_0\exp[-\left(t-D_2\right)^2/T^2]$, where
$\chi_0$ includes the effects of Bose-enhancement~\cite{MAC00}, i.e.,
$\chi_0\propto\sqrt{\rho}$\,. The two-photon (intermediate) detuning is
$\Delta$ ($\delta$), and the mean-field shift due to
atom-atom (atom-molecule, molecule-molecule) interactions is
$\Lambda_{aa}=\rho\lambda_{aa}=4\pi\hbar\rho a_{aa}/m$
($\Lambda_{ag}=\rho\lambda_{gg}=3\pi\hbar\rho a_{ag}/m$,
$\Lambda_{gg}=\rho\lambda_{gg}=2\pi\hbar\rho a_{gg}/m$), where $m$ is
the atom mass and $a_{aa}$ ($a_{ag}$, $a_{gg}$) is the atom-atom
(atom-molecule, molecule-molecule) scattering length.  Spontaneous decay
is included with the rate
$\gamma_s$, which is generally large enough to justify neglect of any
mean-field shifts for the excited-molecular state.

Explicit numbers for
$^{87}$Rb are~\cite{DRU02} $\gamma_s=7.4\times 10^7 \rm{s}^{-1}$,
$\chi_0=2.1\times 10^6\sqrt{\rho/\rho_0}\:\rm{s}^{-1}$,
$\rho_0=4.3\times 10^{14}\,\rm{cm}^{-3}$,
$\lambda_{aa}=4.96\times 10^{-11}\rm{cm}^3/s$,
$\lambda_{ag}=-6.44\times 10^{-11}\rm{cm}^3/s$; although unknown,
the stable-molecule mean-field shift
$\lambda_{gg}=2.48\times 10^{-11}\rm{cm}^3/s$ is estimated by assuming
equal atom-atom and molecule-molecule scattering lengths.

The idea of optimizing the
density of the initial atomic Bose-Einstein condensate is drawn from
investigations into forming a molecular condensate
via Feshbach-resonant interactions\cite{FESH}, i.e., magnetoassociation,
where collision-induced vibrational relaxation of the molecules
is included as a complex particle-particle
scattering length, and where a moderate density helps to alleviate the
associated irreversible losses~\cite{MOD_DEN}.

\begin{figure}
\centering
\includegraphics[width=8.0cm]{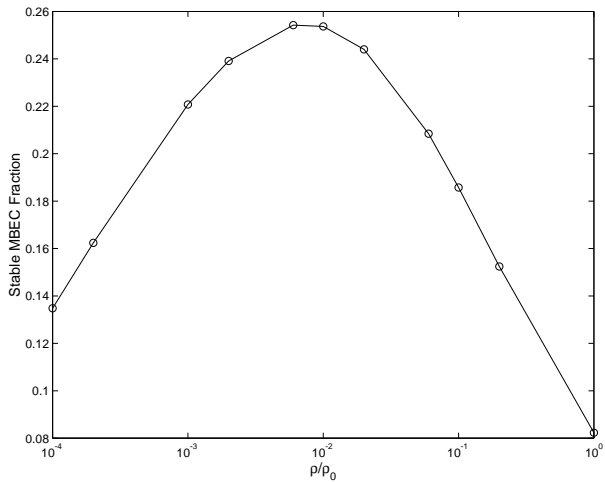}
\caption{Stimulated Raman adiabatic
passage in photoassociation of a freely interacting $^{87}$Rb
Bose-Einstein condensate. For a given $\chi_0(\rho)$, the pulse
parameters are
$\Omega_0=50\chi_0$, $T=5\times 10^3/\chi_0$, $D_1=4.5T$ and $D_2=2.5T$.
The two-photon (intermediate) detuning is
$\Delta=0$ ($\delta=\chi_0$). Note that, since a molecule is born of
two atoms, $|g|^2=1/2$ is actually complete conversion. Mean-field
shifts are marginalized as the density is decreased, but the timescale for
\protect{STIRAP} ($\propto 1/\sqrt{\rho}\,$) also increases, and
spontaneous decay eventually spoils the show.}
\label{BSTIRAP}
\end{figure}

Consider for example the density $\rho=4.3\times
10^{12}\,\rm{cm}^{-1}$, so that
$\chi_0=2.1\times 10^5\rm{s}^{-1}$,
$\Lambda_{aa}=213\,\rm{s}^{-1}$,
$\Lambda_{gg}=107\,\rm{s}^{-1}$,
and $\Lambda_{ag}=277\,\rm{s}^{-1}$. The mean-field shifts are
then roughly three orders of magnitude smaller than the peak Bose-enhanced
free-bound coupling
$\chi_0$, and since $\chi_0$ sets the timescale for
atom-molecule STIRAP~\cite{MAC00}, we expect a smaller role for particle
interactions compared to when $\rho=\rho_0$. This intuition is
confirmed in Fig.~\ref{BSTIRAP}. Note
that efficient short-pulse conversion requires asymmetric
pulses~\cite{HOP01,DRU02}. The optimal density is
$\tilde\rho\sim 10^{12}\,{\rm cm}^{-3}$, for which
conversion to a molecular BEC occurs on a timescale
$T=5\times 10^3/\chi_0 \sim 50\,\rm{ms}$. Optimizing the laser
parameters~\cite{DRU02} should give even further improvement.

Granted, experiments are inhomogeneous and our theory is not. But
inhomogeneity may be approximated by averaging over a
Thomas-Fermi density distribution, and the subsequent optimal density
should be in the same ballpark as the homogeneous prediction.
For $N=5\times 10^5$ atoms in a spherically symmetric trap,
reducing the trap frequency from
$\omega_r/2\pi=100\,\rm{Hz}$~\cite{DRU02} to
$\omega_r/2\pi\sim 1\,\rm{Hz}$
decreases the peak density to roughly the optimal value. If the
photoassociation laser fails to overlap with the expanded BEC for
experimentally reasonable intensities, the cloudsize could be adjusted by
lowering the atom number, or Feshbach-tuning the scattering
length~\cite{FESH}, or a combination thereof. 

One-body correlations are
neglected as per our original work~\cite{MAC00} (see also
Ref.~\cite{NAI03}). Similarly, rogue pair correlations due to
photodissociation to noncondensate modes can be prominent for low
density~\cite{JAV02}; however, the density dependence is weak, and we
show elsewhere~\cite{MAC04} that, as expected, rogue correlations are
safely neglected for low excited-state fractions ($\lsimeq 10^{-6}$) and
photodissociation that is much slower than spontaneous decay.

Additional molecular levels and bound-bound transitions
induced by the photoassociation laser~\cite{DRU02} are also neglected. The
intermediate detunings herein  ($\delta=\chi_0\lsimeq 10^6\,\rm{s}^{-1}$)
are trivial compared to the molecular level spacing ($\sim
10^{10}\,\rm{s}^{-1}$), and the lower density allows
for exact two-photon resonance (compared to $\Delta\neq 0$ in
Ref.~\cite{DRU02}). Omitting these physics is not unreasonable.

Finally, the so-formed molecules are vibrationally
hot~\cite{DRU02}, and molecule-molecule and atom-molecule
collisions may thus foster relaxation to lower-lying vibrational levels.
While the quenching rate is unknown, a lower density is
favorable on this account also (like the mean-field shifts, the quenching
rate is proportional to $\rho$), and a lower peak bound-bound Rabi
coupling ($\Omega_0/\chi_0=50$ compared to $10^3$~\cite{DRU02}) means
that, for user-friendly bound-bound laser intensities, it should be
possible to target vibrationally cooler (i.e., lower-lying) levels. All
told, favorable quenching rates should be achievable.

In conclusion, we suggest that the role of
particle-particle collisions can be further downplayed in STIRAP from
an atomic to a molecular condensate, and the conversion efficiency thereby
improved, by optimizing the initial BEC density along with the
laser parameters. Moreover, we have shown elsewhere that Feshbach-tuning
the scattering length can improve the molecular conversion efficiency
even further~\cite{MAC04}. Hence, if all available knobs are tweaked,
experiments may yet wind up close to the ideal~\cite{MAC00} of near-unit
efficiency.

Support: Academy of
Finland (MM, projects 43336 and 50314); Magnus Ehrnrooth Foundation
(AC); NSF and NASA (JJ, PHY-0097974 and NAG8-1428);
express thanks (from JJ) to Helsinki University of Technology and
Matti Kaivola for support and hospitality.

\end{document}